\documentclass{emulateapj}
\usepackage{natbib}
\newcommand{\Rsun}{\mbox{$\mathrm{R}_{\odot}$}}
\newcommand{\Msun}{\mbox{$M_{\odot}$}}

\begin{document}

\title{High-speed photometry of the disintegrating planetesimals at
  WD\,1145+017: \\ evidence for rapid dynamical evolution}

\author{B. T. G\"ansicke\altaffilmark{1},
        A. Aungwerojwit\altaffilmark{2},
	T.R. Marsh\altaffilmark{1},
        V.S. Dhillon\altaffilmark{3,4},
        D.I. Sahman\altaffilmark{3},
        Dimitri Veras\altaffilmark{1},
        J. Farihi\altaffilmark{5},
        P. Chote\altaffilmark{1},
        R. Ashley\altaffilmark{1},
        S. Arjyotha\altaffilmark{6},
        S. Rattanasoon\altaffilmark{7},
        S.P. Littlefair\altaffilmark{3},
        D. Pollacco\altaffilmark{1},
        M.R. Burleigh\altaffilmark{8}}

\altaffiltext{1}{Department of Physics, University of Warwick,
  Coventry CV4 7AL, UK; boris.gaensicke@warwick.ac.uk}

\altaffiltext{2}{Department of Physics, Faculty of Science, Naresuan
  University, Phitsanulok 65000, Thailand}

\altaffiltext{3}{Department of Physics and Astronomy, University of
  Sheffield, Sheffield S3 7RH, UK}

\altaffiltext{4}{Instituto de Astrof\'sica de Canarias, E-38205 La
  Laguna, Santa Cruz de Tenerife, Spain}
			
\altaffiltext{5}{Department of Physics and Astronomy, University
  College London, London WC1E 6BT, UK}

\altaffiltext{6}{Program of Physics, Faculty of Science and
  Technology, Chiang Rai Rajabhat University, Chiang Rai, 57100,
  Thailand}

\altaffiltext{7}{National Astronomical Research Institute of Thailand,
  191 Siriphanich Bldg., Huay Kaew Rd., Suthep, Muang, Chiang Mai
  50200, Thailand}

\altaffiltext{8}{Department of Physics and Astronomy, University of
  Leicester, Leicester, LE1 7RH, UK} 


\begin{abstract}
We obtained high-speed photometry of the disintegrating planetesimals
orbiting the white dwarf WD\,1145+017, spanning a period of four
weeks. The light curves show a dramatic evolution of the system since
the first observations obtained about seven months ago. Multiple
transit events are detected in every light curve, which have varying
durations ($\simeq3-12$\,min) and depths
($\simeq10-60$\%). The time-averaged extinction is $\simeq11$\,\%,
  much higher than at the time of the \emph{Kepler} observations. The
shortest-duration transits require that the occulting cloud of debris
has a few times the size of the white dwarf, longer events are often
resolved into the superposition of several individual transits. The
transits evolve on time scales of days, both in shape and in depth,
with most of them gradually appearing and disappearing over the course
of the observing campaign. Several transits can be tracked across
multiple nights, all of them recur on periods of $\simeq4.49$\,h,
indicating multiple planetary debris fragments on nearly identical
orbits. Identifying the specific origin of these bodies within this
planetary system, and the evolution leading to their current orbits
remains a challenging problem.
\end{abstract}

\section{Introduction}
Planets around main-sequence stars are ubiquitous
\citep{cassanetal12-1, fressin13-1}, and a significant fraction of
them are predicted to survive the giant branch evolution of their host
stars \citep{mustill+villaver12-1, verasetal13-1}. This expectation is
corroborated by the detection of debris accreted into the photospheres
of white dwarfs, resulting from the tidal disruption \citep{jura03-1,
  debesetal02-1} of planetary bodies among $\simeq25-50$\,\% of all
white dwarfs \citep{zuckermanetal03-1, koesteretal14-1}. Little is
known so far regarding the detailed nature of the disrupted objects,
the exact origin within their planetary systems, and the processes
resulting in their disintegration and subsequent circularization
\citep{debesetal12-2, verasetal14-1, verasetal15-1} and leading to dusty
debris disks with typical radii of $\simeq1\Rsun$, which have been
detected as infrared excess to $\simeq40$ white dwarfs
\citep{rocchettoetal15-1}. The life times of these disks are thought
to be long compared to human time scales, $\sim10^4-10^6$\,yr
\citep{girvenetal12-1}, yet, a small number of disks show substantial
variability on time scales of years to decades in their infrared flux
\citep{xuetal14-2} or in the strength and morphology of optical
emission lines from gaseous disk components \citep{wilsonetal14-1,
  wilsonetal15-1, manseretal15-1}, indicative of ongoing dynamical
processes.

Recently, \citet{vanderburgetal15-1} announced the discovery of
transits in the \emph{K2} light curve of the white dwarf WD\,1145+017,
recurring every $\simeq4.5$\,h, i.e. near the Roche-limit for a
strengthless rubble pile. As WD\,1145+017\footnote{WD\,1145+017 was
  first identified as a white dwarf by \citet{bergetal92-1} in the
  Large Bright Quasar Survey, and later re-discovered as part of the
  Hamburg-ESO survey \citep{friedrichetal00-1}. Adopting a white dwarf
  mass of $0.6\,M_\odot$, \citet{vanderburgetal15-1} determined
  $T_\mathrm{eff}=15\,900$\,K, corresponding to a cooling age of
  175\,Myr, and a distance of 174\,pc.} also exhibits both a large
infrared excess, as well as strong photospheric metal pollution,
\citet{vanderburgetal15-1} interpreted the transit events as the
signature of debris clouds from a disintegrating planetesimal
occulting the white dwarf. Given the presence of multiple
periodicities in the \emph{K2} light curve, \citet{vanderburgetal15-1}
argued for the presence of several, possibly six individual
planetesimals. Following up WD\,1145+017 with ground-based photometry,
\citet{vanderburgetal15-1} and \citet{crolletal15-1} detected several
relatively short ($\simeq10$\,min) and deep ($\simeq40$\,\%) transit
events that did, however, only occasionally repeat at the same phase
during subsequent observations. These optical transits were compatible
with the $\simeq4.5$\,h period identified in the \emph{K2}
data. High-resolution spectroscopy obtained by \citet{xuetal15-1}
reveals strong, broad circumstellar low-ionization absorption lines of
several metals, adding further evidence to the ongoing disintegration
of a planetary body or bodies.

Here we report the first high-speed photometry of the transits at
WD\,1145+017, which show a dramatic evolution of the system in the seven
months since the observations of \citet{vanderburgetal15-1} and
  \citet{crolletal15-1} and suggest a rapid evolution of the planetesimals
and their debris.

\begin{figure*}
\includegraphics[width=2\columnwidth]{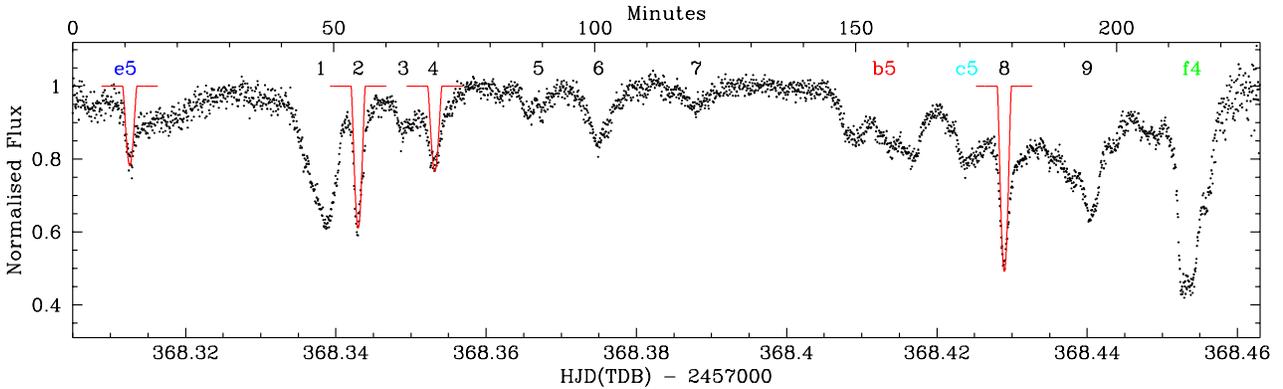}
\caption{\label{f-zoom} TNT/ULTRASPEC high-speed (5\,s) photometry of
  WD\,1145+017, obtained over 3.9\,h on 2015 December 11, illustrating
  the varied and complex nature of the multiple transit events. Many
  of the broader transits display sub-structure that appears to be the
  superposition of several shorter events, e.g. \#4 and b5. Transits
  labeled in color can be tracked in phase across multiple nights (see
  Fig.\,\ref{f-lightcurves}). Simple transit models are overlaid in
  red, see Sect.\,\ref{s-discussion} for details.}
\end{figure*}

\section{Observations}
We obtained high-speed photometry with the frame-transfer camera
ULTRASPEC \citep{dhillonetal14-1} mounted on the 2.4\,m Thai National
Telescope (TNT) on Doi Inthanon over eleven nights in between 2015,
November 28, and December 22. We used a KG5 short-pass filter which
cuts off red light beyond 7000\,\AA, and exposure times of three to
eight seconds, with 15\,ms dead time between exposures. Additional
observations were obtained on 2015, December 17, 23, 24, and 25, using
the Warwick 1\,m (W1m) telescope at the Roque de Los Muchachos
Observatory on La Palma, a robotic F/7 equatorial fork-mounted
telescope with a dual-beam camera system.  These data were acquired in
engineering mode using the reflecting arm of the instrument, which
currently contains a fixed $V+R$ filter with an Andor DW936 camera
(featuring a back-illuminated 2k x 2k CCD with 13.5\,$\mu$m pixels).
The camera was operated with 2 x 2 binning to reduce the readout time
to ~3\,s, giving an exposure cadence of ~23\,s. The combined TNT and
W1m observations add up to 39.3\,h on-target, and were carried out
under a variety of conditions. A full log of the observations is
provided with the online version of the article. Differential
photometry was computed using the nearby comparison stars
SDSS\,J114840.49+012954.7 and SDSS\,J114825.30+013342.2. As we used
non-standard broad-band filters in both instruments, the data were not
absolutely calibrated, and we normalized all light curves to an
out-of-transit flux of one. The normalised light curves will be made
available through the VizieR service at CDS.

\begin{figure*}
\centerline{\includegraphics[width=1.88\columnwidth]{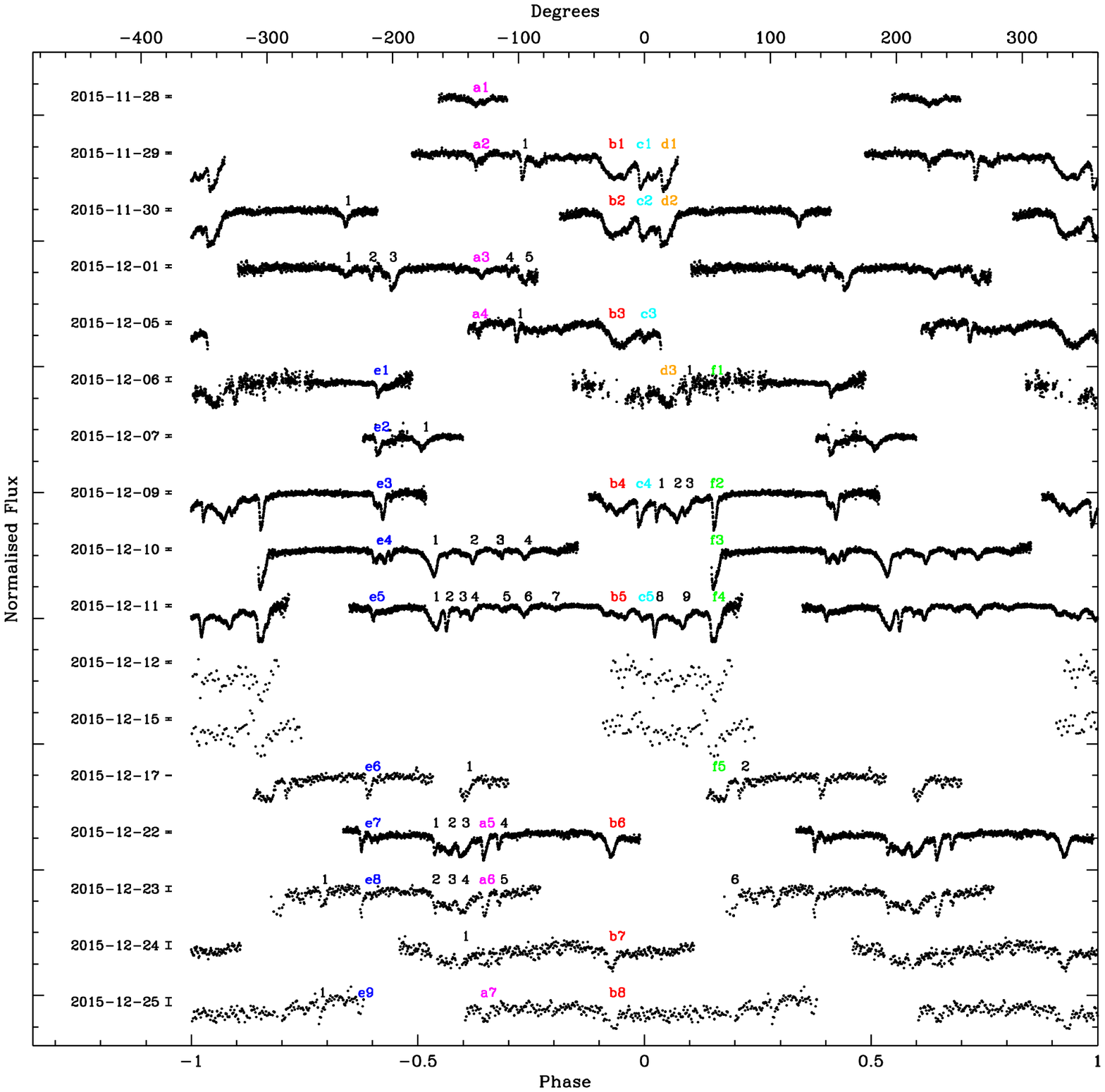}}
\caption{\label{f-lightcurves} TNT and Warwick\,1m light curves of
  WD\,1145+017, folded on a period of 4.4930\,h, two orbital cycles
  are shown for clarity. Typical photometric uncertainties
  are shown on the right to the observing dates. The
    observations on December 6 and 7 were affected by cirrus, and some
    poor quality data was removed. Transit events that can be tracked across
  multiple nights are labeled in color, and were used for a linear
  ephemeris fit (Table\,\ref{t-periods}).}
\end{figure*}

A typical TNT/ULTRASPEC light curve is shown in Fig.\,\ref{f-zoom}. It
is immediately apparent that the morphology of the transits has
dramatically evolved since April/May 2015, when the observations of
\citet{vanderburgetal15-1} and \citet{crolletal15-1} were taken: The
light curve is riddled with numerous transit events varying in
duration from $\simeq3$\,min to $\simeq12$\,min with depths of
$\simeq10$\% to $\simeq60$\%. Many of the transit features overlap,
such that there are only short segments of the light curve which
appear unattenuated by debris.  The full sequence of light curves
obtained during this campaign shown in Fig.\,\ref{f-lightcurves}
(folded on a period of 4.4930\,h, see Sect.\ref{s-ta}) illustrates
that the transits undergo significant night-to-night variations, but
that nevertheless individual events can be confidently identified and
tracked across multiple nights.

\section{Transit Analysis}
\label{s-ta}
The complex and highly variable morphology of the transit events
challenges standard period searches. The strongest signal in an
Analysis-of-Variance (AoV) periodogram is found at 4.498\,h, close to
the period that \citet{vanderburgetal15-1} reported from the analysis
of the \emph{K2} data. However, phase-folding all light curves on this period
washes out many of the features. 

\begin{deluxetable}{lrr}
\tablecolumns{3}
\tablewidth{0pc}
\tablecaption{\label{t-periods} Orbital periods derived from a linear ephemeris
  fit to the six groups of transit events identified in
  Fig.\,\ref{f-lightcurves}. The phase stability of the transits is
  shown in Fig.\,\ref{f-oc}.}
\tablehead{
\colhead{Transits} &
\colhead{Period} &
\colhead{Uncertainty}\\
& \colhead{[h]} &
\colhead{[h]} 
}
\startdata
a & 4.49337 & 0.00021\\
b & 4.49252 & 0.00011\\
c & 4.49257 & 0.00052\\
d & 4.49355 & 0.00040\\
e & 4.49110 & 0.00006\\
f & 4.49513 & 0.00046\\
mean & 4.4930 & 0.0013 
\enddata
\end{deluxetable}

We identified distinct transit features in each light curve, and
measured mid-transit times visually cross-correlating each transit
profile with its mirror image with respect to time. This method has
been developed initially for the analysis of eclipsing cataclysmic
variables \citep{pyrzasetal12-2}, and has been found to be more robust
in the case of asymmetric transit shapes compared to fitting an
analytical function. Estimated uncertainties in the mid-transit times
are $30-60$\,sec, depending on the shape of the individual transit. We
then attempted to group individual transit features that can be
identified across multiple nights, beginning with the broad,
irregularly shaped dip labeled b1--b8 in
Fig.\,\ref{f-lightcurves}. Fitting a linear ephemeris to the transit
times of this feature results in $P=4.49252(11)$\,h, where the
uncertainty is purely statistical in nature. The true uncertainty in
the period is likely to be larger due to the complex shape of the
transit, and the associated difficulty in accurately defining the
mid-transit time.

\begin{figure}
\includegraphics[width=\columnwidth]{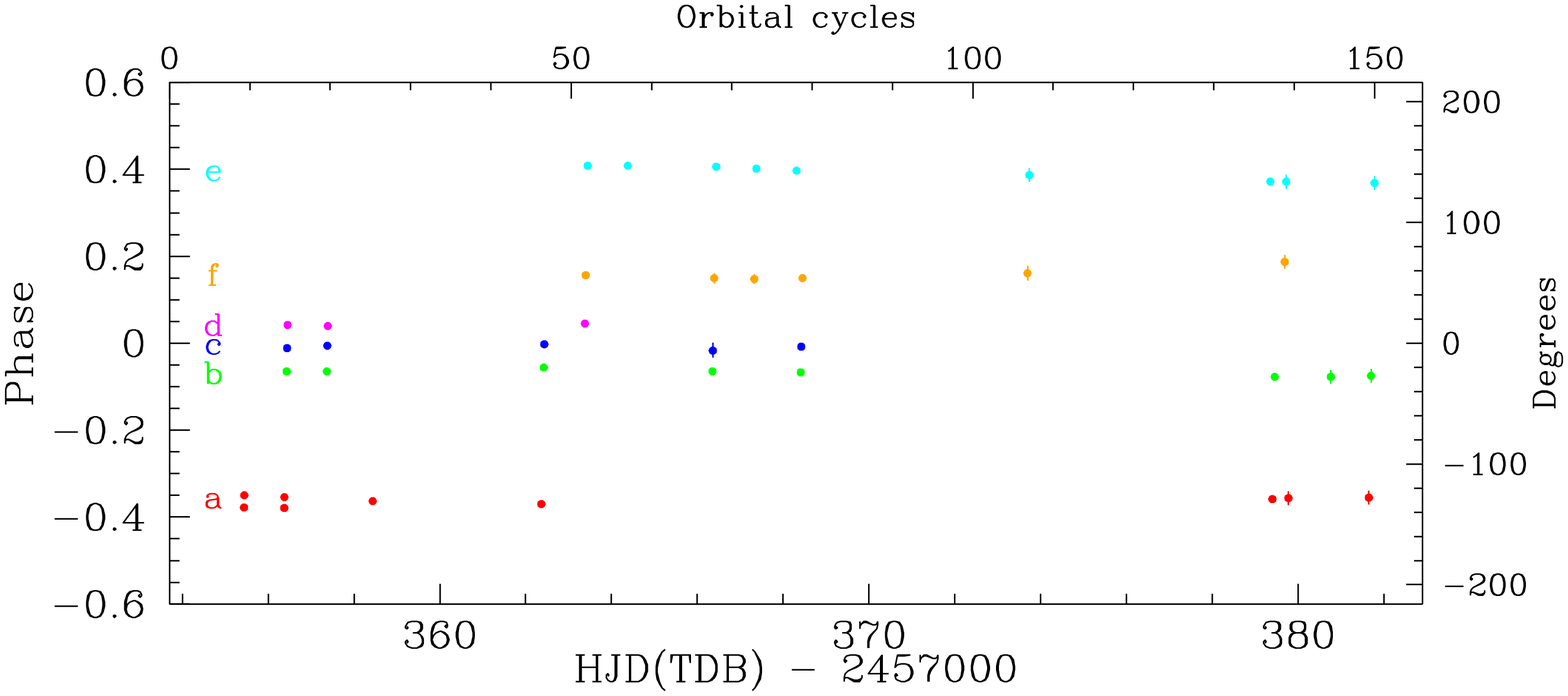}
\caption{\label{f-oc} Relative phases of the six transit events that
  we identified across multiple nights adopting a period of
  4.4930\,h. The color coding of the transits is the same as in
  Fig.\,\ref{f-lightcurves}.}
\end{figure}

We identified five additional groups of transits that appear to be
related to the same material occulting the white dwarf on at
  least three different nights (labeled a, c, d, e, f in
Fig.\,\ref{f-lightcurves}; \#a1 and \#a2 are the superposition of two
very closely spaced short events), and linear ephemeris fits to the
corresponding groups of transit times result in periods in the range
4.491\,h to 4.495\,h (Table\,\ref{t-periods}), with a mean of
$4.4930(13)$\,h. Adopting this mean period, we computed orbital phases
for the six groups of transits, which are shown in Fig.\,\ref{f-oc}.
Various other transit features can be identified in two
  subsequent light curves (e.g. 2015-12-10\#1,2,3,4 and
  2015-12-11\#1,4,5,6; 2015-12-22\#1--4 and 2015-12-23\#2--5), which
  are also consistent with a 4.49\,h period. We were not able to
identify a group of transit events with a significantly different
period, in particular we were not able to recover the periods longer
than $\simeq4.49$\,h discussed by \citet{vanderburgetal15-1}. The
periods from our analysis are similar to those found by
\citet{crolletal15-1}, and are all slightly, but significantly,
shorter than the 4.4988\,h period derived from the \textit{K2} data
\citep{vanderburgetal15-1}. The uncertainties in the periods
  derived for the transit groups a--f are too large to identify them
  with any of the features from \citet{crolletal15-1}.

The light curves, folded on the mean period derived from the linear
ephemeris analysis (Fig.\,\ref{f-lightcurves}) clearly show that the
structure and evolution of the debris orbiting WD\,1145+017 is
extremely complex. Many of the transit events show sub-structure that
changes from night to night (e.g. e2--e4), other strong features
gradually appear (e.g. f2--f5, with possibly a weak pre-cursor,
f1), while others intermittently disappear (e.g. the ``a'' transit
  is not detected on December 10, 11, 17). While we cannot exclude the
  presence of transits recurring on different periods, the stability
  of the relative phases of several groups of transit events suggests
  the presence of many, co-orbital, debris bodies at WD\,1145+017.

\section{Discussion}
\label{s-discussion}
The optical transit events observed by \citet{vanderburgetal15-1} and
\citet{crolletal15-1} had a sharp ingress and a slow egress,
reminiscent to those of the disintegrating planets KIC\,12557548b
\citep{rappaportetal12-1} and KOI\,2700b \citep{rappaportetal14-1}. In
all three cases, the long egress was interpreted as a comet-like tail
trailing the transiting object \citep{vanderburgetal15-1,
  rappaportetal12-1, budaj13-1}.

Most of the transit events in our observations
(Fig.\,\ref{f-lightcurves}) are fairly symmetric, which is a clear
change with respect to the previous observations of WD\,1145+017. One
of the transits, 2015-12-10\#1 and 2014-12-11\#1, exhibits a slow
ingress and sharp egress, possibly suggesting ejection of dust in the
leading direction of the planetesimal. A similar, though much less
pronounced, transit shape has been observed in the \emph{K2} light
curve of the candidate disintegrating planet K2-22b, where it was
modelled with Roche lobe overflow of the planet
\citep{sanchis-ojeda15-1}. The rapid changes seen in the light curve
of WD\,1145+017, compared to the currently known disintegrating
(candidate) planets, KIC\,12557548b, KOI\,2700b and K2-22b, indicates
a much faster evolution of the planetesimals orbiting the white dwarf.

Because of the high time resolution of our data, all individual events
are fully resolved, and it is evident that there is a minimum width of
the transits, $\simeq3$\,min. For comparison with these short transit
features, we calculated the eclipse of the white dwarf by opaque
spheres. The spheres were assumed to be in circular orbits of period
4.49 hours and to have negligible mass compared to the white
dwarf. The mass of the white dwarf was taken to be $0.6\,\Msun$ from
which its radius ($R_\mathrm{wd}=0.0124\,\Rsun$) and the orbital
semi-major axis ($a=1.161\,\Rsun$) follow from the white dwarf
mass-radius relation and Kepler's third law, respectively. The white
dwarf was assumed to be limb-darkened with a linear limb darkening
coefficient of 0.3, leaving the orbital inclination and the radius of
the obscuring sphere as the only free parameters. The flux at any
given phase was calculated by splitting the visible face of the white
dwarf into a regular series of 200 annuli using the Python/C code
developed for the eclipsing double white dwarf CSS\,41177
\citep{boursetal14-1}. In this model, the size of the white dwarf
largely defines a lower limit to the durations of the transits.

The widths of the observed short-duration transits require the
occulting objects to be $2-4$ times the size of the white
dwarf. Synthetic transit profiles for an inclination of $87.75^\circ$
and a transiting object four times bigger than the white dwarf are
superimposed on several of the narrow transit events detected in the
TNT/ULTRASPEC light curve of WD\,1145+017 on 2015 December 11 (scaled
for the different transit depths, being equivalent to different
opacities of the debris clouds). Our admittedly simplistic
  model provides a reasonably good match to the symmetric transit
  profiles, with no need for a trailing tail used to fit the earlier
  observations \citep{vanderburgetal15-1,crolletal15-1}. The large
physical extent of the material causing the transits (several Earth
radii), together with the rapid variability of their shapes and
depths, corroborates the conclusion of \citet{vanderburgetal15-1} that
the transits are not caused by solid bodies, but by clouds of gas and
dust flowing from significantly smaller objects that remain
undetected. The presence of gas along the line-of-sight was verified
by the Keck/HIRES spectra obtained by \citet{xuetal15-1}.  The
  time-averaged extinction throughout the TNT observations is
  $\tau\simeq11$\,\%.  Adopting an absorption coefficient of
  $\kappa(V+B)=1000\,\mathrm{cm^{2}g^{-1}}$ in the visual-blue
  \citep{ossenkopfetal92-1} and a uniform dust distribution in a
  cylindrical sheet of radius $a$ and height $2R_\mathrm{wd}$, we
  estimate  $M_\mathrm{dust}=2\pi
  a\times2R_\mathrm{wd}\,\tau\kappa^{-1}\simeq1.4\times10^{17}$\,g
  to explain the observed extinction. This value should be
  considered as a lower limit, as the vertical extension of the dust is
  likely larger than the white dwarf. The observed rapid variability of the
  transit features suggests that this dust mass could be replenished
  on time scales of days, which would imply a dust production rate of
  $\sim10^{11}\mathrm{g\,s^{-1}}$, similar to the highest accretion
  rates onto metal-polluted white dwarfs \citep{girvenetal12-1}.  The
(presumably rocky) planetesimals may be too small to be detected in
ground-based photometry, or alternatively their orbital inclination
may be such that they do not transit the white dwarf. In the latter
case, the observed transits are likely to be partial, arising from an
expanding gas and dust envelope around the solid bodies.

For a typical white dwarf mass of $\simeq0.6\,\Msun$, an orbital
period of 4.49\,h is close to the tidal disruption radius of a
strengthless rubble-pile with a density of
$\simeq3-4\,\mathrm{g\,cm^{-3}}$ \citep{davidsson99-1, jura03-1,
  verasetal14-1}. The fact that the planetesimals at WD\,1145+017 have
been in these orbits for at least $\simeq550$\,d, or $\simeq3000$
orbital periods, (the \emph{K2} Campaign\,1 data was obtained 2014 June
to August) suggests that they have low eccentricities: tidal
disruption is a strong function of orbital pericentre and almost
independent of semimajor axis \citep{verasetal14-1}, suggesting that
radial incursions due to a nonzero eccentricity would cause disruption
on a timescale much shorter than the observed lifetime of the
planetesimals. If the parent body had significant internal strength,
it could survive within the Roche-radius, allowing for a mild
eccentricity of the orbit.

The origin of the planetesimals on close-in, nearly-circular orbits at
WD\,1145+017 remains challenging to explain, particularly because of
their unknown size.

If the observed signatures represent fragments of just one large
($R\ga1000$\,km) parent body, then a possible scenario for its origin
is the following: (1) within the protoplanetary disc, several planets
formed at or migrated to a distance of several au, (2) during the
giant branch phases of the WD\,1145+017 progenitor, the orbits of
these planets were pushed outward by a factor of a few, (3) after the
star became a white dwarf, the planets underwent a scattering event,
causing one planet to achieve a highly-eccentric orbit with a
pericentre that lay just at the edge of the white dwarf Roche radius,
(4) tidal interactions circularized this orbit, while concurrently the
close proximity to the white dwarf sheared off portions of the planet
and sublimated some of its material.  The first three parts of the
above situation are visualized in Fig. 1 of
\citet{veras+gaensicke15-1}.  The fourth part appears viable because
of the likely similarity to the tidal circularization process of
planets around main sequence stars (see Sec. 5 of
\citealt{veras16-1}). The disruption, and subsequent accretion of a
(minor) planet would result eventually in an extreme metal-pollution
of WD\,1145+017. While the above sequence appears fine-tuned, several
other white dwarfs are known to have accreted at least
$10^{24}-10^{25}$\,g of mainly rocky debris \citep{girvenetal12-1}, equivalent to
the masses of Ceres to Pluto. WD\,1145+017 has already accreted a
minimum of $\simeq6.6\times10^{23}$\,g \citep{xuetal15-1}, which does
not account for the unknown mass still in orbit around it.

If, in contrast, the observations are the consequence of one or more
asteroid-sized ($\sim1$\,km) bodies on nearly-circular orbits, then a
similar scenario as outlined before is unlikely to work because the
asteroid would induce a tidal bulge on the white dwarf that is too
small (e.g. consider a white dwarf-equivalent version of Eqs.\,1--2 of
\citealt{mustill+villaver12-1}) to circularize the orbit.  The
asteroid further would be too large to be circularized by radiation
alone, either through the Yarkovsky effect or Poynting-Robertson drag
\citep{verasetal15-3, verasetal15-1}.  One possibility is that some
fraction of the asteroid has sublimated. Although the impulsive
perturbation from this process would not change the orbital pericentre
\citep{verasetal15-2}, the gas generated at the orbital pericentre
might drag the asteroid into a circular orbit. Circumstellar gas has
been detected at WD\,1145+017 \citep{xuetal15-1}, and gaseous
components of debris disks have been identified at several other white
dwarfs \citep{gaensickeetal06-3, gaensickeetal08-1,
  melisetal12-1}. However, it is unclear if the amount of  gas in
these systems is sufficient to significantly contribute to the
circularisation process. In contrast to comets where out-gassing of
volatiles can free copious quantities of gas, most metal-polluted
white dwarfs are best explained by the accretion of volatile-depleted
asteroids (e.g. \citealt{gaensickeetal12-1, xuetal14-1}, with possibly
a few exceptions, \citealt{farihietal13-2, raddietal15-1}), where gas
production via sublimation is less efficient.

In either case, the time over which we can expect to observe the
breakup of the planetesimal is highly dependent on just how close it
lies to the boundary of the Roche sphere \citep{verasetal14-1}, which
is non-trivially dependent on the unknown physical properties of the
asteroid.  Accompanying the tidal fragmentation might be breakup due
to rotational fission \citep{verasetal14-2}, which is or is not
occurring depending on the previous spin rate of the asteroid as it
settled into its current orbit.  For an older and cooler white dwarf
like WD\,1145+017, fragmentation due to penetration into the Roche
sphere would likely be the dominant breakup mechanism.

\section{Conclusions}
High-speed photometry of WD\,1145+017 obtained over a period of four
weeks reveals frequent transit events with a range of durations, and
reaching depths of up to 60\%. The shortest events last
$\simeq3$\,min, and require an obscuring debris cloud that is a few
times the size of the white dwarf. Longer transits have significant
sub-structure, and often appear to be superpositions of several
individual events. Several groups of transits are identified across
multiple nights and repeat on periods of 4.491 to 4.495\,h, slightly
shorter than the periodicity determined from the \textit{K2} data. The
planetary debris at WD\,1145+017 appears to be undergoing a rapid
evolution, and continued observations are encouraged to determine how
many disintegrating bodies are orbiting this white dwarf, and whether
their nearly co-orbital configuration remains stable over longer
periods of time. Simultaneous multi-color photometry would provide
some insight into the grain size distribution and composition,
\citep[e.g.][]{bochinskietal15-1, crolletal15-1}, and possibly into
changes of these parameters with time, as the system keeps evolving.

\acknowledgements{} We thank the referee for the very prompt
  and constructive report. This work has made use of data obtained at
the Thai National Observatory on Doi Inthanon, operated by NARIT. The
research leading to these results has received funding from the
European Research Council under the European Union's Seventh Framework
Programme (FP/2007-2013) / ERC Grant Agreement n. 320964 (WDTracer).
A.A. acknowledges the supports of the Thailand Research Fund (grant
no. MRG5680152) and the National Research Council of Thailand (grant
no. R2559B034). VSD and TRM acknowledge the support of the Royal
Society and the Leverhulme Trust for the operation of ULTRASPEC at the
TNT. TRM acknowledges support from the Science and Technology
Facilities Council (STFC), grant ST/L000733/1.

\bibliographystyle{apj}
%\bibliography{aamnem99,aabib}

\begin{thebibliography}{}
\expandafter\ifx\csname natexlab\endcsname\relax\def\natexlab#1{#1}\fi

\bibitem[{{Berg} {et~al.}(1992){Berg}, {Wegner}, {Foltz}, {Chaffee}, \&
  {Hewett}}]{bergetal92-1}
{Berg}, C., {Wegner}, G., {Foltz}, C.~B., {Chaffee}, F.~H., J., \& {Hewett},
  P.~C. 1992, ApJS, 78, 409

\bibitem[{{Bochinski} {et~al.}(2015){Bochinski}, {Haswell}, {Marsh}, {Dhillon},
  \& {Littlefair}}]{bochinskietal15-1}
{Bochinski}, J.~J., {Haswell}, C.~A., {Marsh}, T.~R., {Dhillon}, V.~S., \&
  {Littlefair}, S.~P. 2015, ApJ Lett., 800, L21

\bibitem[{{Bours} {et~al.}(2014){Bours}, {Marsh}, {Parsons}, {Copperwheat},
  {Dhillon}, {Littlefair}, {G{\"a}nsicke}, {Gianninas}, \&
  {Tremblay}}]{boursetal14-1}
{Bours}, M.~C.~P., {Marsh}, T.~R., {Parsons}, S.~G., {et~al.} 2014, MNRAS, 438,
  3399

\bibitem[{{Budaj}(2013)}]{budaj13-1}
{Budaj}, J. 2013, A\&A, 557, A72

\bibitem[{{Cassan} {et~al.}(2012){Cassan}, {Kubas}, {Beaulieu}, {Dominik},
  {Horne}, {Greenhill}, {Wambsganss}, {Menzies}, {Williams}, {J{\o}rgensen},
  {Udalski}, {Bennett}, {Albrow}, {Batista}, {Brillant}, {Caldwell}, {Cole},
  {Coutures}, {Cook}, {Dieters}, {Prester}, {Donatowicz}, {Fouqu{\'e}}, {Hill},
  {Kains}, {Kane}, {Marquette}, {Martin}, {Pollard}, {Sahu}, {Vinter},
  {Warren}, {Watson}, {Zub}, {Sumi}, {Szyma{\'n}ski}, {Kubiak}, {Poleski},
  {Soszynski}, {Ulaczyk}, {Pietrzy{\'n}ski}, \& {Wyrzykowski}}]{cassanetal12-1}
{Cassan}, A., {Kubas}, D., {Beaulieu}, J.-P., {et~al.} 2012, Nat, 481, 167

\bibitem[{{Croll} {et~al.}(2015){Croll}, {Dalba}, {Vanderburg}, {Eastman},
  {Rappaport}, {DeVore}, {Bieryla}, {Muirhead}, {Han}, {Latham}, {Beatty},
  {Wittenmyer}, {Wright}, {Johnson}, \& {McCrady}}]{crolletal15-1}
{Croll}, B., {Dalba}, P.~A., {Vanderburg}, A., {et~al.} 2015, ArXiv e-prints,
  arXiv:1510.06434

\bibitem[{{Davidsson}(1999)}]{davidsson99-1}
{Davidsson}, B.~J.~R. 1999, Icarus, 142, 525

\bibitem[{{Debes} {et~al.}(2012){Debes}, {Kilic}, {Faedi}, {Shkolnik},
  {Lopez-Morales}, {Weinberger}, {Slesnick}, \& {West}}]{debesetal12-2}
{Debes}, J.~H., {Kilic}, M., {Faedi}, F., {et~al.} 2012, ApJ, 754, 59

\bibitem[{{Debes} \& {Sigurdsson}(2002)}]{debesetal02-1}
{Debes}, J.~H., \& {Sigurdsson}, S. 2002, ApJ, 572, 556

\bibitem[{{Dhillon} {et~al.}(2014){Dhillon}, {Marsh}, {Atkinson}, {Bezawada},
  {Bours}, {Copperwheat}, {Gamble}, {Hardy}, {Hickman}, {Irawati}, {Ives},
  {Kerry}, {Leckngam}, {Littlefair}, {McLay}, {O'Brien}, {Peacocke},
  {Poshyachinda}, {Richichi}, {Soonthornthum}, \& {Vick}}]{dhillonetal14-1}
{Dhillon}, V.~S., {Marsh}, T.~R., {Atkinson}, D.~C., {et~al.} 2014, MNRAS, 444,
  4009

\bibitem[{{Farihi} {et~al.}(2013){Farihi}, {G{\"a}nsicke}, \&
  {Koester}}]{farihietal13-2}
{Farihi}, J., {G{\"a}nsicke}, B.~T., \& {Koester}, D. 2013, Science, 342, 218

\bibitem[{{Fressin} {et~al.}(2013){Fressin}, {Torres}, {Charbonneau}, {Bryson},
  {Christiansen}, {Dressing}, {Jenkins}, {Walkowicz}, \&
  {Batalha}}]{fressin13-1}
{Fressin}, F., {Torres}, G., {Charbonneau}, D., {et~al.} 2013, ApJ, 766, 81

\bibitem[{{Friedrich} {et~al.}(2000){Friedrich}, {Koester}, {Christlieb},
  {Reimers}, \& {Wisotzki}}]{friedrichetal00-1}
{Friedrich}, S., {Koester}, D., {Christlieb}, N., {Reimers}, D., \& {Wisotzki},
  L. 2000, A\&A, 363, 1040

\bibitem[{{G{\"a}nsicke} {et~al.}(2012){G{\"a}nsicke}, {Koester}, {Farihi},
  {Girven}, {Parsons}, \& {Breedt}}]{gaensickeetal12-1}
{G{\"a}nsicke}, B.~T., {Koester}, D., {Farihi}, J., {et~al.} 2012, MNRAS, 424,
  333

\bibitem[{{G{\"a}nsicke} {et~al.}(2008){G{\"a}nsicke}, {Koester}, {Marsh},
  {Rebassa-Mansergas}, \& {Southworth}}]{gaensickeetal08-1}
{G{\"a}nsicke}, B.~T., {Koester}, D., {Marsh}, T.~R., {Rebassa-Mansergas}, A.,
  \& {Southworth}, J. 2008, MNRAS, 391, L103

\bibitem[{{G{\"a}nsicke} {et~al.}(2006){G{\"a}nsicke}, {Marsh}, {Southworth},
  \& {Rebassa-Mansergas}}]{gaensickeetal06-3}
{G{\"a}nsicke}, B.~T., {Marsh}, T.~R., {Southworth}, J., \&
  {Rebassa-Mansergas}, A. 2006, Science, 314, 1908

\bibitem[{{Girven} {et~al.}(2012){Girven}, {Brinkworth}, {Farihi},
  {G{\"a}nsicke}, {Hoard}, {Marsh}, \& {Koester}}]{girvenetal12-1}
{Girven}, J., {Brinkworth}, C.~S., {Farihi}, J., {et~al.} 2012, ApJ, 749, 154

\bibitem[{{Jura}(2003)}]{jura03-1}
{Jura}, M. 2003, ApJ Lett., 584, L91

\bibitem[{{Koester} {et~al.}(2014){Koester}, {G{\"a}nsicke}, \&
  {Farihi}}]{koesteretal14-1}
{Koester}, D., {G{\"a}nsicke}, B.~T., \& {Farihi}, J. 2014, A\&A, 566, A34

\bibitem[{{Manser} {et~al.}(2016){Manser}, {G{\"a}nsicke}, {Marsh}, {Veras},
  {Koester}, {Breedt}, {Pala}, {Parsons}, \& {Southworth}}]{manseretal15-1}
{Manser}, C.~J., {G{\"a}nsicke}, B.~T., {Marsh}, T.~R., {et~al.} 2016, MNRAS,
  455, 4467

\bibitem[{{Melis} {et~al.}(2012){Melis}, {Dufour}, {Farihi}, {Bochanski},
  {Burgasser}, {Parsons}, {G{\"a}nsicke}, {Koester}, \&
  {Swift}}]{melisetal12-1}
{Melis}, C., {Dufour}, P., {Farihi}, J., {et~al.} 2012, ApJ Lett., 751, L4

\bibitem[{{Mustill} \& {Villaver}(2012)}]{mustill+villaver12-1}
{Mustill}, A.~J., \& {Villaver}, E. 2012, ApJ, 761, 121

\bibitem[{{Ossenkopf} {et~al.}(1992){Ossenkopf}, {Henning}, \&
  {Mathis}}]{ossenkopfetal92-1}
{Ossenkopf}, V., {Henning}, T., \& {Mathis}, J.~S. 1992, A\&A, 261, 567

\bibitem[{{Pyrzas} {et~al.}(2012){Pyrzas}, {G{\"a}nsicke}, {Thorstensen},
  {Aungwerojwit}, {Boyd}, {Brady}, {Casares}, {Hickman}, {Marsh}, {Miller},
  {{\"O}gmen}, {Pietz}, {Poyner}, {Rodr{\'{\i}}guez-Gil}, \&
  {Staels}}]{pyrzasetal12-2}
{Pyrzas}, S., {G{\"a}nsicke}, B.~T., {Thorstensen}, J.~R., {et~al.} 2012, PASP,
  124, 204

\bibitem[{{Raddi} {et~al.}(2015){Raddi}, {G{\"a}nsicke}, {Koester}, {Farihi},
  {Hermes}, {Scaringi}, {Breedt}, \& {Girven}}]{raddietal15-1}
{Raddi}, R., {G{\"a}nsicke}, B.~T., {Koester}, D., {et~al.} 2015, MNRAS, 450,
  2083

\bibitem[{{Rappaport} {et~al.}(2014){Rappaport}, {Barclay}, {DeVore}, {Rowe},
  {Sanchis-Ojeda}, \& {Still}}]{rappaportetal14-1}
{Rappaport}, S., {Barclay}, T., {DeVore}, J., {et~al.} 2014, ApJ, 784, 40

\bibitem[{{Rappaport} {et~al.}(2012){Rappaport}, {Levine}, {Chiang}, {El
  Mellah}, {Jenkins}, {Kalomeni}, {Kite}, {Kotson}, {Nelson},
  {Rousseau-Nepton}, \& {Tran}}]{rappaportetal12-1}
{Rappaport}, S., {Levine}, A., {Chiang}, E., {et~al.} 2012, ApJ, 752, 1

\bibitem[{{Rocchetto} {et~al.}(2015){Rocchetto}, {Farihi}, {G{\"a}nsicke}, \&
  {Bergfors}}]{rocchettoetal15-1}
{Rocchetto}, M., {Farihi}, J., {G{\"a}nsicke}, B.~T., \& {Bergfors}, C. 2015,
  MNRAS, 449, 574

\bibitem[{{Sanchis-Ojeda} {et~al.}(2015){Sanchis-Ojeda}, {Rappaport},
  {Pall{\`e}}, {Delrez}, {DeVore}, {Gandolfi}, {Fukui}, {Ribas}, {Stassun},
  {Albrecht}, {Dai}, {Gaidos}, {Gillon}, {Hirano}, {Holman}, {Howard},
  {Isaacson}, {Jehin}, {Kuzuhara}, {Mann}, {Marcy}, {Miles-P{\'a}ez},
  {Monta{\~n}{\'e}s-Rodr{\'{\i}}guez}, {Murgas}, {Narita}, {Nowak}, {Onitsuka},
  {Paegert}, {Van Eylen}, {Winn}, \& {Yu}}]{sanchis-ojeda15-1}
{Sanchis-Ojeda}, R., {Rappaport}, S., {Pall{\`e}}, E., {et~al.} 2015, ApJ, 812,
  112

\bibitem[{{Vanderburg} {et~al.}(2015){Vanderburg}, {Johnson}, {Rappaport},
  {Bieryla}, {Irwin}, {Lewis}, {Kipping}, {Brown}, {Dufour}, {Ciardi}, {Angus},
  {Schaefer}, {Latham}, {Charbonneau}, {Beichman}, {Eastman}, {McCrady},
  {Wittenmyer}, \& {Wright}}]{vanderburgetal15-1}
{Vanderburg}, A., {Johnson}, J.~A., {Rappaport}, S., {et~al.} 2015, Nat, 526,
  546

\bibitem[{{Veras}(2016)}]{veras16-1}
{Veras}, D. 2016, Royal Society Open Science

\bibitem[{{Veras} {et~al.}(2015{\natexlab{a}}){Veras}, {Eggl}, \&
  {G{\"a}nsicke}}]{verasetal15-3}
{Veras}, D., {Eggl}, S., \& {G{\"a}nsicke}, B.~T. 2015{\natexlab{a}}, MNRAS,
  451, 2814

\bibitem[{{Veras} {et~al.}(2015{\natexlab{b}}){Veras}, {Eggl}, \&
  {G{\"a}nsicke}}]{verasetal15-2}
---. 2015{\natexlab{b}}, MNRAS, 452, 1945

\bibitem[{{Veras} \& {G{\"a}nsicke}(2015)}]{veras+gaensicke15-1}
{Veras}, D., \& {G{\"a}nsicke}, B.~T. 2015, MNRAS, 447, 1049

\bibitem[{{Veras} {et~al.}(2014{\natexlab{a}}){Veras}, {Jacobson}, \&
  {G{\"a}nsicke}}]{verasetal14-2}
{Veras}, D., {Jacobson}, S.~A., \& {G{\"a}nsicke}, B.~T. 2014{\natexlab{a}},
  MNRAS, 445, 2794

\bibitem[{{Veras} {et~al.}(2014{\natexlab{b}}){Veras}, {Leinhardt}, {Bonsor},
  \& {G{\"a}nsicke}}]{verasetal14-1}
{Veras}, D., {Leinhardt}, Z.~M., {Bonsor}, A., \& {G{\"a}nsicke}, B.~T.
  2014{\natexlab{b}}, MNRAS, 445, 2244

\bibitem[{{Veras} {et~al.}(2015{\natexlab{c}}){Veras}, {Leinhardt}, {Eggl}, \&
  {G{\"a}nsicke}}]{verasetal15-1}
{Veras}, D., {Leinhardt}, Z.~M., {Eggl}, S., \& {G{\"a}nsicke}, B.~T.
  2015{\natexlab{c}}, MNRAS, 451, 3453

\bibitem[{{Veras} {et~al.}(2013){Veras}, {Mustill}, {Bonsor}, \&
  {Wyatt}}]{verasetal13-1}
{Veras}, D., {Mustill}, A.~J., {Bonsor}, A., \& {Wyatt}, M.~C. 2013, MNRAS,
  431, 1686

\bibitem[{{Wilson} {et~al.}(2014){Wilson}, {G{\"a}nsicke}, {Koester}, {Raddi},
  {Breedt}, {Southworth}, \& {Parsons}}]{wilsonetal14-1}
{Wilson}, D.~J., {G{\"a}nsicke}, B.~T., {Koester}, D., {et~al.} 2014, MNRAS,
  445, 1878

\bibitem[{{Wilson} {et~al.}(2015){Wilson}, {G{\"a}nsicke}, {Koester}, {Toloza},
  {Pala}, {Breedt}, \& {Parsons}}]{wilsonetal15-1}
---. 2015, MNRAS, 451, 3237

\bibitem[{{Xu} \& {Jura}(2014)}]{xuetal14-2}
{Xu}, S., \& {Jura}, M. 2014, ApJ Lett., 792, L39

\bibitem[{{Xu} {et~al.}(2015){Xu}, {Jura}, {Dufour}, \&
  {Zuckerman}}]{xuetal15-1}
{Xu}, S., {Jura}, M., {Dufour}, P., \& {Zuckerman}, B. 2015, ArXiv e-prints,
  arXiv:1511.05973

\bibitem[{{Xu} {et~al.}(2014){Xu}, {Jura}, {Koester}, {Klein}, \&
  {Zuckerman}}]{xuetal14-1}
{Xu}, S., {Jura}, M., {Koester}, D., {Klein}, B., \& {Zuckerman}, B. 2014, ApJ,
  783, 79

\bibitem[{{Zuckerman} {et~al.}(2003){Zuckerman}, {Koester}, {Reid}, \&
  {H{\"u}nsch}}]{zuckermanetal03-1}
{Zuckerman}, B., {Koester}, D., {Reid}, I.~N., \& {H{\"u}nsch}, M. 2003, ApJ,
  596, 477

\end{thebibliography}

\end{document}